\documentclass[12pt]{article}
\usepackage{amsmath}
\usepackage{amssymb}
\usepackage{graphicx}
\usepackage{axodraw}
\usepackage{epsfig}
\usepackage{url}
\usepackage{fancybox}
\usepackage[small]{caption}
\begin{document}
\baselineskip 0.6cm

\def\simgt{\mathrel{\lower2.5pt\vbox{\lineskip=0pt\baselineskip=0pt
           \hbox{$>$}\hbox{$\sim$}}}}
\def\simlt{\mathrel{\lower2.5pt\vbox{\lineskip=0pt\baselineskip=0pt
           \hbox{$<$}\hbox{$\sim$}}}}
\def\simprop{\mathrel{\lower3.0pt\vbox{\lineskip=1.0pt\baselineskip=0pt
             \hbox{$\propto$}\hbox{$\sim$}}}}
\def\bra#1{\left< #1 \right|}
\def\ket#1{\left| #1 \right>}
\def\inner#1#2{\left< #1 | #2 \right>}
\def\chairup{\begin{picture}(15,15)
  \Line(4,10)(11,10) \Line(4,4)(11,4) \Line(4,4)(4,10) \Line(11,4)(11,10)
  \Line(4,4)(2.5,1) \Line(11,4)(12.5,1) \Line(2.5,1)(12.5,1)
  \Line(3,1)(3,-5) \Line(12,1)(12,-5)
  \Line(4.5,1)(4.5,-2.5) \Line(10.5,1)(10.5,-2.5)
\end{picture}}
\def\chairdown{\begin{picture}(15,15)
  \Line(4,-5)(11,-5) \Line(4,1)(11,1) \Line(4,1)(4,-5) \Line(11,1)(11,-5)
  \Line(4,1)(2.5,4) \Line(11,1)(12.5,4) \Line(2.5,4)(12.5,4)
  \Line(3,4)(3,10) \Line(12,4)(12,10)
  \Line(4.5,4)(4.5,7.5) \Line(10.5,4)(10.5,7.5)
\end{picture}}
\def\ftchairup{\begin{picture}(12,12)
  \Line(3.2,8)(8.8,8.0) \Line(3.2,3.2)(8.8,3.2)
  \Line(3.2,3.2)(3.2,8.0) \Line(8.8,3.2)(8.8,8.0)
  \Line(3.2,3.2)(2.0,0.8) \Line(8.8,3.2)(10.0,0.8) \Line(2.0,0.8)(10.0,0.8)
  \Line(2.4,0.8)(2.4,-4.0) \Line(9.6,0.8)(9.6,-4.0)
  \Line(3.6,0.8)(3.6,-2.0) \Line(8.4,0.8)(8.4,-2.0)
\end{picture}}
\def\ftchairdown{\begin{picture}(12,12)
  \Line(3.2,-4.0)(8.8,-4.0) \Line(3.2,0.8)(8.8,0.8)
  \Line(3.2,0.8)(3.2,-4.0) \Line(8.8,0.8)(8.8,-4.0)
  \Line(3.2,0.8)(2.0,3.2) \Line(8.8,0.8)(10.0,3.2) \Line(2.0,3.2)(10.0,3.2)
  \Line(2.4,3.2)(2.4,8.0) \Line(9.6,3.2)(9.6,8.0)
  \Line(3.6,3.2)(3.6,6.0) \Line(8.4,3.2)(8.4,6.0)
\end{picture}}
\def\observer{\begin{picture}(15,15)
  \CArc(7.5,7.5)(2.5,0,360)
  \put(5.2,4.9){$\cdot$} \Line(5.0,7.8)(4.6,6.6) \Line(6.6,6.2)(5.5,6.0)
  \Line(7.5,4.5)(3,1.5) \Line(7.5,4.5)(12,1.5) \Line(7.5,4.5)(7.5,0.5)
  \Line(7.5,0.5)(5,-5) \Line(7.5,0.5)(10,-5)
\end{picture}}
\def\brainup{\begin{picture}(15,15)
  \Line(-3,9)(11,9) \CArc(11,7)(2,0,90) \Line(13,7)(13,-0.2)
  \CArc(-3,5)(4,0,90) \Line(1,5)(1,-0.2)
  \CArc(3,-0.2)(2,180,270) \Line(3,-2.2)(11,-2.2) \CArc(11,-0.2)(2,270,360)
  \Line(5,7)(8.5,7) \Line(5,4)(8.5,4)
  \Line(5,4)(5,7) \Line(8.5,4)(8.5,7)
  \Line(5,4)(4.25,2.5) \Line(8.5,4)(9.25,2.5) \Line(4.25,2.5)(9.25,2.5)
  \Line(4.5,2.5)(4.5,-0.5) \Line(9,2.5)(9,-0.5)
  \Line(5.25,2.5)(5.25,0.75) \Line(8.25,2.5)(8.25,0.75)
\end{picture}}
\def\braindown{\begin{picture}(15,15)
  \Line(-3,9)(11,9) \CArc(11,7)(2,0,90) \Line(13,7)(13,-0.2)
  \CArc(-3,5)(4,0,90) \Line(1,5)(1,-0.2)
  \CArc(3,-0.2)(2,180,270) \Line(3,-2.2)(11,-2.2) \CArc(11,-0.2)(2,270,360)
  \Line(5,-0.5)(8.5,-0.5) \Line(5,2.5)(8.5,2.5)
  \Line(5,2.5)(5,-0.5) \Line(8.5,2.5)(8.5,-0.5)
  \Line(5,2.5)(4.25,4) \Line(8.5,2.5)(9.25,4) \Line(4.25,4)(9.25,4)
  \Line(4.5,4)(4.5,7) \Line(9,4)(9,7)
  \Line(5.25,4)(5.25,5.75) \Line(8.25,4)(8.25,5.75)
\end{picture}}

\begin{titlepage}

\begin{flushright}
UCB-PTH-12/12 \\
\end{flushright}

\vskip 1.3cm

\begin{center}
{\Large \bf Complementarity Endures:\\
\vspace{1mm}
No Firewall for an Infalling Observer}

\vskip 0.7cm

{\large Yasunori Nomura, Jaime Varela, and Sean J. Weinberg}

\vskip 0.4cm

{\it Berkeley Center for Theoretical Physics, Department of Physics,\\
 University of California, Berkeley, CA 94720, USA}

\vskip 0.1cm

{\it Theoretical Physics Group, Lawrence Berkeley National Laboratory,
 CA 94720, USA}

\vskip 0.8cm

\abstract{We argue that the complementarity picture, as interpreted as a 
 reference frame change represented in quantum gravitational Hilbert space, 
 does not suffer from the ``firewall paradox'' recently discussed by Almheiri, 
 Marolf, Polchinski, and Sully.  A quantum state described by a distant 
 observer evolves unitarily, with the evolution law well approximated by 
 semi-classical field equations in the region away from the (stretched) 
 horizon.  And yet, a classical infalling observer does not see a violation 
 of the equivalence principle, and thus a firewall, at the horizon. 
 The resolution of the paradox lies in careful considerations on how 
 a (semi-)classical world arises in unitary quantum mechanics describing 
 the whole universe/multiverse.}

\end{center}
\end{titlepage}

\section{Introduction: Complementarity and Firewall}
\label{sec:intro}

In the past decades, it has become increasingly clear that the concept of 
spacetime must receive substantial revisions when it is treated in a fully 
quantum mechanical manner.   Consider describing a process in which an 
object falls into a black hole, which eventually evaporates, from the 
viewpoint of a distant observer.  Unitarity of quantum mechanics, together 
with the semi-classical analysis of the process, suggests that the information 
content of the object will first be stored on the (stretched) horizon, 
and then emitted back to distant space in the form of quantum correlations 
among Hawking radiation quanta.  On the other hand, the equivalence 
principle implies that the object should not find anything special at 
the horizon, when the process is described by an observer falling with 
the object.  These two pictures are obviously incompatible if we formulate 
quantum mechanics in global spacetime; in particular, the full quantum 
information content of the falling object will be duplicated into internal 
spacetime and the horizon/Hawking radiation, violating the no-cloning 
theorem of quantum mechanics~\cite{Wootters:1982zz}.

Black hole complementarity~\cite{Susskind:1993if,Stephens:1993an} asserts 
that there is no contradiction between the two pictures, since the 
statements by the two observers cannot be operationally compared 
due to the nontrivial structure of the geometry.  Specifically, 
no physical observer can collect the information from both Hawking 
radiation and the fallen object, avoiding a violation of the no-cloning 
theorem~\cite{Hayden:2007cs,Susskind:1993mu}.  This requires a deviation 
from the naive expectation that global spacetime provides a universal platform 
on which unitary quantum mechanics is formulated preserving (approximate) 
locality.  Extensions of the complementarity picture to other spacetimes 
are discussed in Refs.~\cite{Banks:2001yp,Dyson:2002nt,Susskind:2002ri,%
Nomura:2011dt,Nomura:2011rb}.  In particular, it has profound implications 
on how the eternally inflating universe/multiverse must be described without 
leading to many apparent problems/paradoxes~\cite{Nomura:2011dt,Nomura:2012zb}.

Recently, the complementarity picture has been challenged by Almheiri, 
Marolf, Polchinski, and Sully (AMPS)~\cite{Almheiri:2012rt}.  These authors 
consider a black hole that forms from collapse of some pure state and 
discuss what an observer falling into this black hole after the Page time 
(the time when the black hole emits a half of its initial Bekenstein-Hawking 
entropy) will see.  AMPS argue that the observer measures high energy 
quanta at the horizon, a phenomenon referred to as a ``firewall,'' 
contradicting the equivalence principle.  The argument (which we will 
review in more detail in the next section) goes as follows:\ because 
of the entanglement between early radiation and other degrees of freedom, 
implied by the purity of a state, an infalling observer may select 
a state that is {\it in}compatible with the equivalence principle by 
making measurements on early radiation, which can be causally accessed 
by the observer.  This puzzling conclusion is called the firewall 
paradox.  In a subsequent paper~\cite{Susskind:2012ey} Susskind has 
argued that the location and formation time scale of the firewall are 
different from what AMPS imagine, but it still requires a violation 
of the equivalence principle in a low-curvature region.  More recently, 
Bousso has refuted these arguments by considering that each causal 
diamond has its ``own theory''~\cite{Bousso:2012as}.  This, however, 
still leaves the question of why the infalling observer does not select 
a state that violates the equivalence principle by measuring early 
radiation.

In this paper, we argue that the firewall paradox can be resolved by 
considering carefully how a semi-classical world appears in a fully 
quantum mechanical description of the entire universe/multiverse.  In 
Section~\ref{sec:resol}, we review the argument by AMPS and discuss 
how the paradox can be avoided by treating quantum measurements as 
self-consistent dynamical processes occurring in unitary quantum 
mechanics.  In Section~\ref{sec:compl}, we consider the issue in the 
context of an explicit realization of complementarity described in 
Ref.~\cite{Nomura:2011rb}, which applies to arbitrary dynamical spacetimes 
including cosmological ones.  In this framework, complementarity is 
understood as a reference frame change represented in quantum gravitational 
Hilbert space, which corresponds to only limited regions in spacetime 
associated with a fixed reference frame.  This implies that complementarity, 
as interpreted in Ref.~\cite{Nomura:2011rb}, is a relation between full 
quantum states as viewed from different reference frames, and not a 
relation of different views in a fixed semi-classical background. 
We argue that the framework does not suffer from the firewall paradox. 
In Section~\ref{sec:concl} we conclude:\ the complementarity picture 
endures in the fully quantum mechanical context.

\section{Resolution of the Paradox---Quantum Measurement and the 
 Emergence of a Classical World}
\label{sec:resol}

Following AMPS, let us consider a black hole that has formed from collapse 
of some pure state and has already emitted more than a half of its initial 
Bekenstein-Hawking entropy in the form of Hawking radiation (a black hole 
that is older than the Page time $t_{\rm Page} \approx O(M^3)$, where $M$ 
is the mass of the original black hole).  Because of the purity of the 
state, the system can be written as
\begin{equation}
  \ket{\Psi} = \sum_i c_i \ket{\psi_i} \otimes \ket{i},
\label{eq:state}
\end{equation}
where $\ket{\psi_i} \in {\cal H}_{\rm rad}$ represent states associated 
with Hawking radiation emitted early on and thus macroscopically 
away from the black hole, while $\ket{i} \in {\cal H}_{\rm BH}$ the 
degrees of freedom associated with the region near the horizon.  Because 
the black hole is old, the dimensions of the Hilbert space factors 
${\cal H}_{\rm rad}$ and ${\cal H}_{\rm BH}$ satisfy ${\rm dim}\,{\cal 
H}_{\rm rad} \gg {\rm dim}\,{\cal H}_{\rm BH}$~\cite{Page:1993wv}.  Then, 
as argued by AMPS, states $\ket{\psi_i}$ for different $i$ are expected to 
be nearly orthogonal, so that one can construct a projection operator $P_i$ 
that acts only on ${\cal H}_{\rm rad}$ ({\it not} on ${\cal H}_{\rm BH}$) 
but selects a particular term in Eq.~(\ref{eq:state}) corresponding 
to a specific state $\ket{i}$ in ${\cal H}_{\rm BH}$ when operated 
on $\ket{\Psi}$:
\begin{equation}
  P_i \ket{\Psi} \propto \ket{\psi_i} \otimes \ket{i}.
\label{eq:projection}
\end{equation}
(Note that there is no summation on $i$ here.)  The point is that one 
can construct such an operator for an arbitrary state $\ket{i}$ in 
${\cal H}_{\rm BH}$.

Now, consider a distant observer who is located far away from the black 
hole horizon and an infalling observer who enters into the black hole 
after the Page time is passed.  These two observers can both access 
early Hawking radiation, i.e.\ elements in ${\cal H}_{\rm rad}$, causally. 
From the viewpoint of the distant observer, the remaining degrees of 
freedom in ${\cal H}_{\rm BH}$ are those located near/on the stretched 
horizon that later evolve into Hawking radiation quanta emitted into 
distant space.  For the infalling observer, on the other hand, the 
degrees of freedom in ${\cal H}_{\rm BH}$ are those associated with the 
spacetime region near the horizon which he/she is traveling through.%
\footnote{Following AMPS (and some other literature), here we take the 
 ``Heisenberg picture'' in which both the distant and infalling observers 
 access the same Hilbert space factor ${\cal H}_{\rm BH}$; these 
 observers interpret it differently because the operator sets they 
 use to extract physical effects on other systems, e.g.\ a freely 
 falling object, are different (which must be the case because of 
 relative acceleration between the two observers).  On the other hand, 
 in Refs.~\cite{Nomura:2011dt,Nomura:2011rb}, the ``Schr\"{o}dinger 
 picture'' has been employed; in this picture, the operator sets 
 associated with the two observers are the same, but the Hilbert space 
 factors are different: ${\cal H}_{\rm BH}$ and ${\cal H}'_{\rm BH}$. 
 These two pictures are equivalent---in the Heisenberg picture, changing 
 the viewpoint corresponds to mapping the operator set associated 
 with one observer to that associated with the other, while in the 
 Schr\"{o}dinger picture, it corresponds to mapping elements of one 
 Hilbert space factor to those in the other.  For more discussions on 
 this and related points, see Section~\ref{sec:compl}. \label{ft:pictures}}

The essence of the argument by AMPS is that since the infalling observer 
can access the early radiation, he/she can select a particular term in 
Eq.~(\ref{eq:state}) by making a measurement on those degrees of freedom 
(due to the ability of constructing an operator $P_i$ corresponding to an 
arbitrary $\ket{i}$).  In particular, AMPS imagine that such a measurement 
would select a term in which $\ket{i}$ in ${\cal H}_{\rm BH}$ is an 
eigenstate of the number operator $b^\dagger b$ of a Hawking radiation 
mode as viewed from a distant observer, which we denote generically by 
$\ket{\tilde{\imath}}$ ($\in {\cal H}_{\rm BH}$):
\begin{equation}
  b^\dagger b \ket{\tilde{\imath}} \propto \ket{\tilde{\imath}}.
\label{eq:i-tilde}
\end{equation}
If this were true, then the infalling observer must find physics 
represented by $\ket{\tilde{\imath}}$ near the horizon, and since 
an eigenstate of $b^\dagger b$ cannot be a vacuum for the infalling 
modes $a_\omega$, related to $b$ by
\begin{equation}
  b = \int_0^\infty\! d\omega 
    \left( B(\omega) a_\omega + C(\omega) a_\omega^\dagger \right)
\label{eq:b_vs_a}
\end{equation}
with some functions $B(\omega)$ and $C(\omega)$, the infalling observer 
must experience nontrivial physics at the horizon (i.e.\ $a_\omega 
\ket{\tilde{\imath}} \neq 0$ for infalling modes with the frequencies 
much larger than the inverse horizon size).  This obviously contradicts 
the equivalence principle.

What can be wrong with this argument?  The point is that any measurement 
that leads to a classical world is {\it a dynamical process} dictated by 
unitary evolution of a state, and {\it not} something we can impose from 
outside by acting with some projection operator on the state.  (A detailed 
explanation of this point in the context of dynamical spacetimes/cosmology, 
see Ref.~\cite{Nomura:2011rb}.)  In particular, {\it the existence of 
the projection operator $P_i$ for an arbitrary $i$ does not imply that a 
measurement---in a sense that it leads to a classical world---can occur 
to pick up the corresponding state $\ket{i}$}.  This point can be understood 
relatively easily if we consider a state corresponding to a superposition 
of two different macroscopic configurations, for example that of upward 
and downward chairs (relative to some other object, e.g.\ the ground, 
which we omit):
\begin{equation}
  \ket{\Psi_{\rm chair}} = \bigl| \chairup \bigr> + \bigl| \chairdown \bigr>.
\label{eq:chair}
\end{equation}
An observer interacting with this system evolves following the unitary, 
deterministic Schr\"{o}dinger equation; in particular, the combined chair 
and observer state becomes
\begin{equation}
  \ket{\Psi_{\rm chair + observer}} 
  = \bigl| \chairup \bigr> \otimes \bigl| \observer\brainup \bigr> 
    + \bigl| \chairdown \bigr> \otimes \bigl| \observer\braindown \bigr>.
\label{eq:chair+obs}
\end{equation}
This does {\it not} lead to a classical world in which the chair is in 
a superposition state but to two different worlds in which the chair 
is upward and downward, respectively.%
\footnote{Note that a classical world can be defined as the world in which 
 the information/observation is stable/reproducible~\cite{q-Darwinism-1,%
 q-Darwinism-2}.  In the example here, the information that the chair 
 is upward or downward is reproduced---or ``amplified''---in many physical 
 systems, e.g.\ in the chair itself, the brain state of the observer, in 
 the conversation he/she has about the chair, etc.  This, however, cannot 
 be the case for the information that the chair is in a superposition 
 state because of the property of the dynamics.}
Namely, the measurement is performed in the particular basis $\bigl\{ 
\bigl| \chairup \bigr>, \bigl| \chairdown \bigr> \bigr\}$, which is 
{\it determined by the dynamics}---the existence of an operator projecting 
onto a superposition chair state does not mean that a measurement can 
be performed in that basis.  For a sufficiently macroscopic object, 
the appropriate basis for measurements is almost always the one in 
which the object has well-defined configurations in classical phase 
space (within some errors, which must exist because of the uncertainty 
principle).  This is because the Hamiltonian has the form that is local 
in spacetime~\cite{Nomura:2011rb}.

In the specific context of the firewall argument, a measurement of the early 
radiation by an infalling observer will select a state in ${\cal H}_{\rm rad}$ 
that represents a well-defined classical configuration of Hawking radiation 
quanta $\ket{\psi_I}$ (which will be entangled with other classical objects, 
e.g.\ a measuring device monitoring the quanta).  The point is that 
{\it there is no reason that the state $\ket{\psi_I}$ selected in this 
way is a state that is maximally entangled with $\ket{\tilde{\imath}}$, 
i.e.\ $\ket{\psi_{\tilde{\imath}}}$}.  Indeed, since the states that 
represent well-defined classical worlds are very special among all 
the possible states%
\footnote{In the example of a chair, those are the states having 
 $\alpha \approx 0$ or $\pi/2$ when we write a general chair state 
 as $\ket{\Psi_{\rm chair}} = \cos\alpha\, \bigl| \ftchairup \bigr> 
 + e^{i\varphi} \sin\alpha\, \bigl| \ftchairdown \bigr>$.}
and since the dimension of the Hilbert space factor spanned by $b^\dagger 
b$ is very small (${\rm dim}\,{\cal H}_{\rm BH} \ll {\rm dim}\,{\cal 
H}_{\rm rad}$), we expect that $\ket{\psi_I}$ does not coincide with 
$\ket{\psi_{\tilde{\imath}}}$ for any $\tilde{\imath}$; see the Appendix 
for a sample calculation demonstrating this.  Namely, the basis in 
${\cal H}_{\rm rad}$ selected by entanglement with the eigenstates 
of $b^\dagger b$ is different from the one selected by entanglement 
with the infalling observer, i.e.\ by measurements.  This implies that 
the state projected by $P_{\tilde{\imath}}$, considered by AMPS, does 
{\it not} correspond to a classical world for the infalling observer, 
but rather to a superposition of macroscopically different worlds.  And 
since the equivalence principle is a statement about a (semi-)classical 
world, there is no contradiction between the complementarity picture and 
the equivalence principle as envisioned by AMPS.  In other words, in 
a classical world for the infalling observer, in which the equivalence 
principle holds, the state in ${\cal H}_{\rm rad}$ is always in 
a superposition of $\ket{\psi_{\tilde{\imath}}}$'s for different 
$\tilde{\imath}$'s, so that the corresponding state in ${\cal H}_{\rm BH}$ 
is {\it not} an eigenstate of $b^\dagger b$---in particular, there is no 
contradiction if it is a simultaneous eigenstate of $a_\omega$'s with 
the eigenvalue zero.

In fact, in order to make the above argument, we need not consider 
a physical observer who is actually falling into the black hole. 
The Hawking radiation quanta in the early radiation has a natural basis 
$\ket{\Psi_I}$ in which the information is amplified, i.e.\ the basis 
corresponding to (emergent) classical worlds; and if we focus on one 
of these terms, then the conclusion by AMPS is avoided.  A state projected 
by $P_{\tilde{\imath}}$ corresponds to a superposition of different 
macroscopic worlds, which is not the one described by classical general 
relativity.  Conversely, a state corresponding to a well-defined classical 
world is one in which the state in ${\cal H}_{\rm rad}$ is a superposition 
of different $\ket{\psi_{\tilde{\imath}}}$'s, so that the corresponding 
state in ${\cal H}_{\rm BH}$ is not an eigenstate of $b^\dagger b$.

The picture described here can also be made explicit in the context 
of the framework in Ref.~\cite{Nomura:2011rb}.  In this framework, 
complementarity is realized for general dynamical spacetimes as a 
reference frame change represented in quantum gravitational Hilbert 
space ${\cal H}_{\rm QG}$, whose elements correspond only to limited 
spacetime regions as viewed from a fixed (local Lorentz) reference 
frame.  We now turn to discussions on this picture.

\section{Complementarity as a Reference Frame Change}
\label{sec:compl}

The complementarity picture refers to the fact that assigning physical 
degrees of freedom in all the global spacetime regions as naively suggested 
by quantum field theory is overcounting.  These degrees of freedom must 
be divided by gauge redundancies associated with a consistent quantum 
theory of gravity, which are expected to be much larger than those in 
general relativity.  The reduction of degrees of freedom appearing in 
a black hole geometry discussed so far is expected to be one manifestation 
of this more general phenomenon.

In the treatment in the previous section, the reduction has been realized 
in such a way that the horizon degrees of freedom as viewed from the 
distant observer and the degrees of freedom located near the horizon 
as viewed from the infalling observer are represented by the same Hilbert 
space factor ${\cal H}_{\rm BH}$.  These two observers interpret states 
in ${\cal H}_{\rm BH}$ differently because the operator sets used to 
extract physical implications, e.g.\ on objects carried by them, differ 
for the two observers because of relative acceleration between them. 
Equivalently, we may formulate the same phenomenon such that the two 
observers use the same set of operators ${\cal O}_X$ but the Hilbert 
space factors they access are different: ${\cal H}_{\rm BH}$ and 
${\cal H}'_{\rm BH}$, whose elements represent only {\it limited}, 
and {\it different} spacetime regions (see also footnote~\ref{ft:pictures}). 
This is the picture adopted in Refs.~\cite{Nomura:2011dt,Nomura:2011rb}, 
and corresponds to describing all the phenomena from the viewpoint of 
a fixed, freely falling---or local Lorentz---reference frame (if one 
chooses the fixed operator set ${\cal O}_X$ to be that of the inertial 
frame).

In this picture, complementarity between physics described by two 
observers---or rather, in two reference frames---is nothing but a unitary 
transformation represented in the full Hilbert space ${\cal H}_{\rm QG}$ 
that contains both ${\cal H}_{\rm BH}$ and ${\cal H}'_{\rm BH}$. 
In particular, building on earlier progresses in quantum gravity, 
such as holography~\cite{'tHooft:1993gx,Susskind:1994vu,Bousso:1999xy} 
and complementarity~\cite{Susskind:1993if,Stephens:1993an}, 
Ref.~\cite{Nomura:2011rb} has proposed a particular structure for 
the Hilbert space for quantum gravity ${\cal H}_{\rm QG}$.  According 
to this proposal, ${\cal H}_{\rm QG}$ contains a component ${\cal H}$ 
that represents general spacetimes as well as ${\cal H}_{\rm sing}$ that 
represents spacetime singularities: ${\cal H}_{\rm QG} = {\cal H} \oplus 
{\cal H}_{\rm sing}$. (The ${\cal H}_{\rm sing}$ component is needed to 
preserve unitarity of the evolution of a generic state.)  In particular, 
elements in ${\cal H}$ correspond to states defined on the past light 
cone of a fixed reference point (the ``spatial'' origin of the reference 
frame) in and on the apparent horizon.  This interpretation of states 
in spacetime arises through responses of these states to the action of 
operators ${\cal O}_X$, and the use of the past light cone, although 
not mandatory, allows us to formulate the theory (the algebra among 
${\cal O}_X$'s) independently of the background spacetime.  (More 
details of this construction will be discussed elsewhere~\cite{NVW}.) 

\begin{figure}[t]
\begin{center}
  \includegraphics[width=5cm]{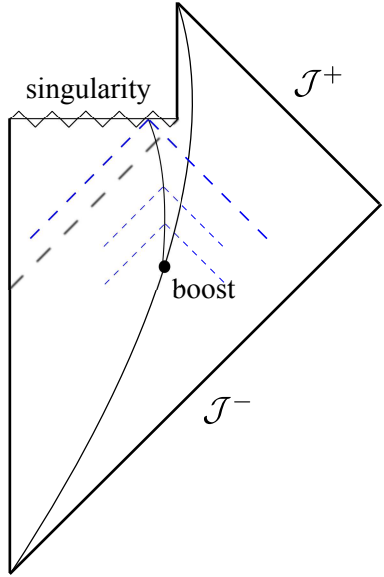}
\caption{A Penrose diagram showing the trajectories of the reference point 
 $p$ for two reference frames---distant (right curve) and infalling (left 
 curve)---in a black hole background.  The two frames can be related 
 by a boost acted at some time, represented by the dot on which the two 
 curves merge.  The dashed hats attached to the left curve illustrate 
 (null) hypersurfaces on which states, as viewed from the infalling frame, 
 are defined.  Here, we represented the setup in a fixed classical geometry 
 for presentation purposes, but a full quantum state is in general a 
 superposition of terms representing well-defined classical geometries.}
\label{fig:BH}
\end{center}
\end{figure}
The framework of Ref.~\cite{Nomura:2011rb} is sufficiently general and 
explicit that we can describe the picture advocated in the previous section 
in a more concrete form.  To be specific, let us consider a process in 
which a physical (test) object falls into an old black hole after the 
Page time.  Suppose we describe this process using a reference frame 
whose spatial origin $p$ stays outside the horizon throughout the entire 
history of the black hole, until it completely evaporates.  (Such a 
trajectory of $p$ is depicted in Fig.~\ref{fig:BH}.)  Then for a generic 
initial state $\ket{\Psi_0}$, the state describing the system evolves as
\begin{equation}
  \ket{\Psi_0} \,\,\stackrel{\rm object\,\,falls}{\longrightarrow}\,\, 
  \ket{\Psi} = \sum_i \ket{\psi_i} \otimes \ket{i}
  \,\,\stackrel{\rm BH\,\,evaporates}{\longrightarrow}\,\, \ket{\Psi_\infty},
\label{eq:distant-gen}
\end{equation}
where $\ket{\psi_i} \in {\cal H}_{\rm rad}$, $\ket{i} \in {\cal H}_{\rm BH}$, 
and $\ket{\Psi_\infty}$ is a final state after the black hole evaporates.%
\footnote{Evolution of a state is possible despite the fact that time 
 translation is a gauge transformation, since the state has a boundary 
 (the apparent horizon) so that it may not be an eigenstate of the 
 Hamiltonian.  This is analogous to the situation in flat space, where 
 the contribution from surface terms~\cite{Arnowitt:1962hi} allows a state 
 not to vanish under the action of the Hamiltonian: $H \ket{\psi} \neq 0$. 
 (This is the reason why the $S$-matrix can ever be discussed in string 
 theory in the flat background.)  For more discussions on this point, 
 see Ref.~\cite{Nomura:2012zb}.}
In this description, once the test object falls into the black hole, it 
is absorbed into the (stretched) horizon, which will eventually give the 
information about the object back in late Hawking radiation.

Now, imagine that the falling object is entangled with a particular term 
in $\ket{\Psi}$ in Eq.~(\ref{eq:distant-gen}), with the corresponding 
$\ket{\psi_i}$ representing a well-defined configuration of radiation 
quanta in classical phase space.  Given the structure of the Hamiltonian, 
i.e.\ locality, this is a perfectly legitimate case to consider.  The 
relevant part of the evolution is then
\begin{equation}
  \ket{\Psi_0} \,\,\stackrel{\rm object\,\,falls}{\longrightarrow}\,\, 
  \ket{\Psi} = \ket{\psi_i} \otimes \ket{i}
  \,\,\stackrel{\rm BH\,\,evaporates}{\longrightarrow}\,\, \ket{\Psi_\infty},
\label{eq:distant}
\end{equation}
where we have kept only the term of the form $\ket{\psi_i} \otimes \ket{i}$ 
in $\ket{\Psi}$, without a summation on $i$.  The picture discussed in the 
previous section then says that $\ket{i}$ in $\ket{\Psi}$ does {\it not} 
evolve into an eigenstate of the number operator $b^\dagger b$ for a late 
Hawking radiation mode.  From the viewpoint of the distant observer, this 
simply says that the final state $\ket{\Psi_\infty}$ in Eq.~(\ref{eq:distant}) 
is a superposition of different worlds in which the configuration of 
late Hawking radiation differ.

What happens if we describe the classical world/history represented 
in Eq.~(\ref{eq:distant}) from a reference frame whose spatial origin 
$p$ falls into the black hole together with the test object?  The 
resulting description can be obtained by first boosting the entire 
state (or equivalently the reference frame) at the initial moment, 
$\ket{\Psi_0} \rightarrow \ket{\Psi'_0} = U_{\rm boost} \ket{\Psi_0}$, 
and then following the evolution of that state to the future (see 
Fig.~\ref{fig:BH} for a schematic depiction of this procedure):
\begin{equation}
  \ket{\Psi'_0} \,\,\stackrel{\rm object\,\,falls}{\longrightarrow}\,\, 
  \ket{\Psi'} = \ket{\psi'_i} \otimes \ket{i'}
  \,\,\stackrel{p\,\,{\rm hits\,\,singularity}}{\longrightarrow}\,\,
  \ket{\Psi'_\infty},
\label{eq:infalling}
\end{equation}
where $\ket{\psi'_i} \in {\cal H}_{\rm rad}$ (i.e.\ $\ket{\psi_i}$ as 
viewed from the new reference frame), $\ket{i'} \in {\cal H}'_{\rm BH}$, 
and $\ket{\Psi'_\infty} \in {\cal H}_{\rm sing}$.  Since $\ket{i}$ in 
Eq.~(\ref{eq:distant}) is not an eigenstate of $b^\dagger b$, there 
is no contradiction if $\ket{i'}$ is a simultaneous zero-eigenvalue 
eigenstate of the annihilation operators $a_\omega$'s in the infalling 
frame: $a_\omega \ket{i'} = 0$, as suggested by the equivalence principle. 
The complementarity picture endures.

The setup considered by AMPS corresponds to projecting $\ket{\Psi}$ 
in Eq.~(\ref{eq:distant-gen}) onto a state that is an eigenstate 
$\ket{\tilde{\imath}}$ of $b^\dagger b$:
\begin{equation}
  \ket{\Psi} \rightarrow P_{\tilde{\imath}} \ket{\Psi} 
  = \ket{\psi_{\tilde{\imath}}} \otimes \ket{\tilde{\imath}}.
\label{eq:proj-Psi}
\end{equation}
What is this state?  To answer this question, note that the argument in 
the Appendix implies that $\ket{\psi_{\tilde{\imath}}}$ is a superposition 
of terms representing different classical worlds:
\begin{equation}
  \ket{\psi_{\tilde{\imath}}} = \sum_a d_a \ket{\psi_a},
\label{eq:psi-tilde-i}
\end{equation}
where $\ket{\psi_a} \in {\cal H}_{\rm rad}$ represent states having 
well-defined classical configurations of emitted Hawking quanta, which 
will be maximally entangled with other classical objects.  This state, 
therefore, is not the one described by general relativity, which is 
a theory for a classical world, i.e.\ a basis state in which the 
information is naturally amplified by the local Hamiltonian.%
\footnote{The argument here implies that one cannot treat a 
 carefully-crafted quantum device that can collect the information 
 in the Hawking radiation quanta to learn that they are in a 
 $\ket{\psi_{\tilde{\imath}}}$ state and send it to the infalling 
 observer, within the context of a classical description of physics. 
 Since the information is encoded in Hawking quanta in a highly scrambled 
 form, such a device would have to be very large collecting many quanta 
 spread in space without losing their coherence, and would be in a 
 superposition of different classical configurations as indicated by 
 Eq.~(\ref{eq:psi-tilde-i}).  Given that ${\rm dim}\,{\cal H}_{\rm rad} 
 \gg {\rm dim}\,{\cal H}_{\rm BH}$, this requires an exponentially 
 fine-tuned (and intrinsically quantum mechanical) initial condition 
 for the device state, and physics occurring to the infalling observer 
 entangled with such a device cannot be well described by general 
 relativity, which is a theory of a classical world.}

Finally, we point out an interesting aspect of the black hole evolution 
represented in Eq.~(\ref{eq:distant}).  To discuss it, let us regard 
emission of Hawking radiation quanta by the black hole as a stochastic 
process in which the black hole performs a random walk in momentum space 
due to backreaction of the Hawking emission.  In a timescale of order 
the Page time, the black hole emits order $N_H \sim M/T_H \sim M^2$ 
Hawking quanta, where $T_H \sim 1/M$ is the Hawking temperature.  This 
implies that each emission occurs with the average time interval of order 
$t_{\rm Page}/N_H \sim M$, with each emission providing $\varDelta v 
\sim \varDelta p/M \sim T_H/M \sim 1/M^2$ kick in velocity space, leading 
to $|v| \sim \varDelta v \sqrt{t_{\rm Page}/M} \sim 1/M$ at $t \sim 
t_{\rm Page}$, whose direction stays almost constant in each kick. 
The location of the black hole, therefore, has large uncertainty of 
order $|v|\, t_{\rm Page} \sim M^2 \gg r_S$ after the Page time, giving 
huge variations of the geometries.%
\footnote{This large uncertainty of the black hole location was noted 
 earlier in Ref.~\cite{Page:1979tc}.}
More detailed discussion of the implications of this large uncertainty 
will be given in a forthcoming paper.

\section{Conclusions}
\label{sec:concl}

In this paper we have argued that the complementarity picture endures 
despite the ``firewall challenge'' recently posed by AMPS.  The point 
is that states corresponding to well-defined classical worlds are very 
special ones in quantum mechanics, which are determined by the dynamics 
as a result of amplification of information, i.e.\ by the evolution of 
the system with an environment (such as a measuring device, observer, 
etc.).  A vast majority of the states in general Hilbert space, including 
the ones considered by AMPS, are superpositions of different classical 
worlds.  And since general relativity is a theory describing a classical 
world, results obtained in it need not persist in those general 
superposition states.

Another issue we have discussed is that complementarity, as interpreted 
in Ref.~\cite{Nomura:2011rb}, is a relation between full quantum states, 
which can involve superpositions of classical spacetimes, as viewed 
from different reference frames---it is not a relation of different 
views in a fixed semi-classical background, as might have been imagined 
in some earlier work.  Further implications of this fact for physics of 
black holes will be discussed in a forthcoming paper.

\vspace{0.3cm}

\begin{flushleft}
{\bf Note added:}

While completing this paper, we received a paper by Harlow~\cite{Harlow} 
which also discusses the firewall paradox.
\end{flushleft}

\section*{Acknowledgments}

We would like to thank Steve Giddings and Joe Polchinski for correspondence 
from which many of the clarifications in the revised version have resulted. 
This work was supported in part by the Director, Office of Science, Office 
of High Energy and Nuclear Physics, of the US Department of Energy under 
Contract DE-AC02-05CH11231, and in part by the National Science Foundation 
under grant PHY-0855653.

\appendix

\section{Misalignment between the Classical State Basis and the Basis 
 Selected by {\boldmath $b^\dagger b$}}
\label{app:calc}

In this appendix we argue that if the black hole is sufficiently 
old, there is an exponential suppression of the probability that 
$\ket{\psi_{\tilde{\imath}}}$, corresponding to {\it some} $b^\dagger b$ 
eigenstate $\ket{\tilde{\imath}}$, represents a well-defined classical 
world.

Let $N_{\rm R} \equiv {\rm dim}\,{\cal H}_{\rm rad}$ and $N_{\rm BH} \equiv 
{\rm dim}\,{\cal H}_{\rm BH}$, and consider the expansion of a pure state 
in Hilbert space ${\cal H}_{\rm rad} \otimes {\cal H}_{\rm BH}$ in the form
\begin{equation}
  \ket{\Psi} = \sum_{\tilde{\imath}=1}^{N_{\rm BH}} c_{\tilde{\imath}} 
    \ket{\psi_{\tilde{\imath}}} \otimes \ket{\tilde{\imath}},
\label{eq:itilde-expans}
\end{equation}
where $\ket{\tilde{\imath}}$ represent the $b^\dagger b$ eigenstates.
We assume that ${\cal H}_{\rm rad}$ has an orthonormal ``classical state 
basis'' $\left\{ e_n \left|\right. n = 1,\ldots,N_{\rm R} \right\}$, 
where $e_n$ represent states that have well-defined configurations 
in classical phase space (within some errors).  In practice, this basis 
is determined by the interaction between the early Hawking radiation 
and (classical) environment.

We now estimate the probability that there exists an $\tilde{\imath}$ 
such that the corresponding $\ket{\psi_{\tilde{\imath}}}$ is ``close'' 
to one of the vectors $\ket{e_n}$.  To proceed, it is necessary to 
introduce the (arbitrary) notion of a state being ``almost 
classical.''  To this end, we introduce a small positive real number 
$\epsilon$ and say that $\ket{\psi_{\tilde{\imath}}}$ is close to 
$\ket{e_n}$ if $|\inner{e_n}{\psi_{\tilde{\imath}}}| > 1 - \epsilon$.

Since ${\cal H}_{\rm rad}$ is an $N_{\rm R}$-dimensional complex vector 
space, we can identify it with a $2N_{\rm R}$-dimensional real vector 
space. Choose a fixed $\ket{\psi_{\tilde{\jmath}}}$ and $\ket{e_m}$, and 
rotate the coordinates such that $\ket{e_m}$ is the $2N_{\rm R}$-dimensional 
vector $(1,0,\cdots,0)$.  The vector 
$\ket{\psi_{\tilde{\jmath}}}$ can then be identified as a normalized 
vector in the $2N_{\rm R}$-dimensional space i.e.\ a point on the 
($2N_{\rm R} - 1$)-dimensional unit sphere, $S^{2N_{\rm R}-1}$.  We 
wish to compute the probability, $P(\tilde{\jmath},m)$, that the condition 
$|\inner{e_m}{\psi_{\tilde{\jmath}}}| > 1 - \epsilon$ is satisfied. 
Assuming that $\ket{\psi_{\tilde{\jmath}}}$ is equally likely to lie 
on any point on the sphere, $P(\tilde{\jmath},m)$ is the twice the 
area of the spherical cap
\begin{equation}
  C = \left\{ (x_1,\ldots,x_{2N_{\rm R}}) \in S^{2N_{\rm R}-1} 
    \left|\right. x_1 > 1-\epsilon \right\},
\label{eq:cap}
\end{equation}
divided by the area of $S^{2N_{\rm R}-1}$ for normalization.  (The factor 
of 2 arises because we must consider both the $x_1 > 1 - \epsilon$ and 
the $x_1 < -(1-\epsilon)$ caps.)  Since $\epsilon$ is small, $C$ is 
approximately a solid ball of radius $\sqrt{2\epsilon}$.  Carrying out 
this calculation and using the fact that $N_{\rm R} \gg 1$, one finds 
that $P(\tilde{\jmath},m) \sim (2\epsilon)^{N_{\rm R}}/\sqrt{N_{\rm R}}$.

Finally, the probability that any of the $\ket{\psi_{\tilde{\imath}}}$'s 
is close to any of the $\ket{e_n}$'s is $P = N_{\rm R} N_{\rm BH} 
P(\tilde{\jmath},m)$.  (The factor of $N_{\rm R}$ arises because there 
are $N_{\rm R}$ classical basis vectors and the factor of $N_{\rm BH}$ 
because there are $N_{\rm BH}$ distinct $\ket{\psi_{\tilde{\imath}}}$.) 
We thus conclude that
\begin{equation}
  P \sim \sqrt{N_{\rm R}} N_{\rm BH} (2\epsilon)^{N_{\rm R}}.
\label{eq:prob}
\end{equation}
The probability that any of the $\ket{\psi_{\tilde{\imath}}}$'s represents 
a classical world is exponentially suppressed in $N_{\rm R}$ for a 
sufficiently old black hole.

\end{document}